# Weak-antilocalization and Surface Dominated Transport in Topological Insulator $Bi_2Se_2Te$

Radha Krishna Gopal[1], Sourabh Singh[1], Ramesh Chandra[2], Chiranjib Mitra[1]

[1]Indian Institute of Science Education Research Kolkata
Mohanpur- 741246, India.

[2]Indian Institute of Technology Roorkee, Roorkee-247667, India.

## Abstract

We explore the phase coherence of thin films of the topological insulator material $Bi_2Se_2Te$ grown through pulsed laser deposition (PLD) technique. The films were characterised using various techniques for phase and composition. The films were found to be of good quality. We carried out extensive magneto-transport studies of these films and found that they exhibit two dimensional weak antilocalization behaviour. A careful analysis revealed a relatively high phase coherence length (58nm at 1.78K) for a PLD grown film. Since PLD is an inexpensive technique, with the possibility to integrate with other materials, one can make devices which can be extremely useful for low power spintronics and topological quantum computation.

# I. Introduction

Topological insulators (TI) are a new state of quantum matter characterized by spin momentum locked gapless surface states which are protected by time reversal symmetry[1–3]. These class of materials were predicted theoretically[4–8] and subsequently experimentally realized a few years ago in HgTe quantum wells which is a 2D TI and bismuth based materials $Bi_2Se_3$, $Bi_2Te_3$ and $Sb_2Te_3$ which are 3D TI by separate groups[9–12]. Ideally topological insulators have insulating bulk band gap and spin polarized conducting electronic surface states (or edge states in two- dimension) arising out of strong spin orbit interaction[13]. These electronic surface states exhibit a relativistic energy – momentum dispersion relation, constituting a single Dirac cone at the Fermi surface[8]. This is a hallmark for a topological insulator, which is characterised by an odd number of Dirac cones at the Fermi surface, as opposed to an analogous material, Graphene, which has an even number of Dirac cones[11,14]. These electronic states are metallic and robust to the time reversal invariant perturbations like non-magnetic impurities, defects and vacancies in the sample and have been experimentally probed by various experimental techniques, such as angle resolved photoemission spectroscopy (ARPES), scanning tunnelling microscopy (STM) and quantum interference experiments at low temperatures [9,11,12,15–17]. In recent times, intense research work has been carried out from a fundamental point of view both in theory as well as experiments to explore the new and exotic properties that topological insulators exhibit owing to their unique phase of quantum matter[18,19]. In addition to this, owing to the topologically protected surface states, these systems also open new platforms for device applications (low power spintronics) and quantum computing[3].

These surface states can also be probed by magnetotransport measurements, and the topologically protected surface states are expected to show weak anti- localization behaviour due to their spin being locked to the momentum, which has remained a characteristic classification tool for the spin orbit coupled systems[20,21]. Numerous works on magneto-transport measurements have proved the existence of surface states by means of weak antilocalization and observation of Shubnikov de Has (SdH) oscillations in the single crystals and thin films of $Bi_2Te_3$, $Bi_2Se_3$ and their ternary chalcogenides[22,23].

In the second generation binary alloys there has always been a problem of unintentional doping which result in dominant bulk charge carriers which overwhelm the surface state conduction[2] It is possible to control the Fermi surface by gating thereby pushing it in the bulk band gap. Apart from gating, doping these alloys with Ca, Sb and Sn has proved to be useful in tuning the chemical potential in the bulk band gap but these doping in turn create defects, thereby rendering it impossible to realize an intrinsic TI[24,25]. We believe that an intrinsic TI state can be created by alloying two TI materials $Bi_2Te_3$ and $Bi_2Se_3$ to form either $Bi_2Se_2Te$ (BST), $Bi_2Te_2Se$ (BTS) or some intermediate ternary tetradymide alloys[25–27]. We were interested in the intermediate ternary compound BST since there has been limited study on it despite BST being predicted as an ideal candidate for intrinsic TI like its sister compound BTS[28]. Furthermore, it has been theoretically proposed that BST has a large spin texture with a Dirac point in the bulk band gap[23,26]. A large spin texture promises a non-trivial Berry phase that restrains the backscattering.

In this study, we present magnetotransport studies of BST thin films, one of the recently discovered ternary tetradymides based topological insulators ($M_2X_2Y$ where M= Bi; X and Y = Se or Te). To the best of our knowledge this is the first time BST thin films have been fabricated using PLD. This material exhibits property very similar to its parent compounds in terms of single Dirac cone but differ in terms of chemical potential which lies in the bulk band gap in these compounds thereby making them better TI candidates[26]. Magnetotransport data (conductance as a function of field) obtained at different temperatures for these thin films exhibits archetypal weak antilocalization (WAL) character in the low temperature regime which is a hallmark of topologically protected surface states with an odd number of Dirac cones and hence signifies a surface dominant transport of charge carriers[29,30].

## II. Sample Preparation and Characterization

BST thin films were grown using pulsed laser deposition (PLD) technique in a flowing Argon atmosphere. Pulsed laser deposition (PLD) has the advantage of maintaining stoichiometry of the target composition in the thin films since it is a non-equilibrium process and is capable of forming samples which may not otherwise form through conventional routes including single crystal growth techniques[31]. This is especially useful since we are depositing a multi-element compound such as BST and for PLD grown films the stoichiometry is preserved during deposition in addition to maintaining a high deposition rate in comparison to other similar techniques like sputtering or molecular beam epitaxy. PLD grown films also have an added advantage that apart from being inexpensive it offers a unique opportunity of depositing multilayers of TI with ready integration with superconductor or magnetic materials for devices applications.

We deposited the thin films using a KrF Excimer Pulsed Laser having a wavelength of 248 nm and a pulsed width of 25 ns. A high purity BST target of 1 inch was used to ablate and the deposition was carried out at a low repetition rate of 1Hz to improve the quality of thin films for transport measurements[32]. The target material for depositing BST thin films was prepared by mixing stoichiometric proportions of $Bi_2Se_3$ and $Bi_2Te_3$ (in 2:1 ratio) and compressing the mixture at high pressure (12 Ton)[25]. The deposition chamber had been evacuated to a reasonably high vacuum with the base vacuum in the chamber being $2\times10^{-6}$ mbar. The films were deposited in a flowing argon atmosphere and a partial pressure of 0.56 mbar was maintained throughout the process. Despite the fact that PLD technique is useful in deposition over small areas in comparison to sputtering, it gives a very high deposition rate and is an efficient technique for the deposition of thin films with multi-element stoichiometry such as BST.

Our films were deposited on silicon (Si 100) substrates. Before deposition silicon substrates were properly cleaned in acetone followed by isopropyl alcohol in ultrasonic bath for 30 minutes. Thin films were grown at different experimental conditions such as partial pressure, substrate temperature, and target to substrate distance, varying energy density at the target and the various parameters were then optimised. The best quality films were fomed when the substrate temperature was maintained at 300 °C. It was observed that the films deposited at a high repetition rate were rough as determined from atomic force microscopy (AFM) technique and were less crystalline in nature as is evident from x-ray diffraction (XRD) data.

The deposition parameters were optimised by investigating the X-ray diffraction, scanning electron microscopy (SEM) images, energy dispersive spectroscopy (EDX) and Raman spectroscopy measurements of the as grown thin films. Thickness of these thin films were determined by surface profilometer and depending on the growth conditions, the typical thickness varied from 100nm to 122nm which is equivalent to 100 to 122 quintuple layers (QL). The films that are being reported here are of the order of 120nm.

To determine the chemical composition and crystalline nature of the thin films we performed Raman spectroscopy, EDX and X- ray diffraction (XRD) measurements. XRD pattern of a typical thin film is shown in Fig. 1. One can clearly see that the films grow along the c-axis. Moreover XRD analysis reveals the disordered occupation of Te/Se atoms in the quintuple layer unit unlike BTS which has ordered occupation of Te/Se atoms in its skippenite crystal structure and due to this fact it has highest possible bulk resistivity found in any topological insulator candidate till now[23, 28]. But despite its deviation from skippenite crystal structure BST has shown remarkable surface state transport properties, outnumbering residual conductance by nearly 57%[23]. BST and BTS both have the same space group ($R\bar{3}m$) as their parent alloys $Bi_2Se_3$ and $Bi_2Te_3$. Since both $Bi_2Se_3$ and $Bi_2Te_3$ parent alloys are strongly Raman active, it is expected that BST will also be Raman active and exhibit its characteristic Raman peaks. The Raman spectrum of BST thin film is shown in Fig. 2. In the low wave number regime the parent compounds show their characteristic $A_{1g}^1$, $E_{2g}$ and $A_{1g}^2$ Raman peaks at positions 116.6 cm$^{-1}$, 146.52 cm$^{-1}$ and 165.13 cm$^{-1}$ which are consistent with the previous reports[33,34].

### III. Results and Discussions

Transport measurements on thin films were carried out in a cryogen free magnet (Cryogenic, U.K.) in a temperature range of 1.78K to 300K and a maximum magnetic field of 8 Tesla. All the magnetotransport measurements were performed on these samples in Van der Pauw geometry. A four probe technique was employed to study all the electrical and magnetotransport properties and for the convenience of transport measurement we deposited thin films of BST on the substrates of dimensions 4mm x 4mm.

Fig. 3 shows the variation of longitudinal resistance with temperature for thin film of BST. The Film exhibits metallic character down to 15K owing to reduced electron phonon interaction but below this the resistance shows an upturn. The possible reason for this metallic behaviour may be the slight stoichiometric mismatch which alters the carrier density. It has been shown for similar samples that the carrier density in these systems is highly sensitive to the Se and Te ratio[23,26]. This upturn in resistance indicates an insulating character due to freezing out of bulk carriers as reported by others in different TI samples. One of the possible reasons suggested for the slight upturn in the R vs T plot is the presence of Coulomb interaction among the electrons in the low temperature regime in the presence of disorder[35]. We do not attribute the low temperature behaviour of resistance to the increased electron-electron (e-e) interaction because up to now it has been observed only in ultra-thin films and in that case topological protection of carriers goes away and the two surfaces of TI thin film couple with each other driving it in to the insulating regime at low temperature[29] and thickness of our samples was of the order of 120nm. Rather we ascribe it to the depleted bulk carriers and weak localization of trivial two dimensional electron gas (2DEG) in presence of disorder. There are clear evidences of 2DEG coexisting with topologically protected Dirac states on TI thin films hence it is a competition between WAL due to topologically protected surface states and Weak Localization (WL) due to 2DEG in presence of disorder which determines the low temperature behaviour of thin films, in parallel freezing of bulk carriers[36–38]. There is no conclusive evidence of e-e interaction accounting for this slight upturn in R-T behaviour and it requires a systematic study both experimentally and theoretically. Further improvements are in process in terms of sample fabrication of BST to make it more insulating in the bulk in the whole temperature regime and will be reported elsewhere.

From the Hall measurements a typical carrier density of the order of $9 \times 10^{19} cm^{-3}$ was calculated at 1.78K and from the sign of Hall coefficient type of charge carrier was found to be electron. Thin films deposited with PLD technique show a low carrier mobility, and we obtained a mobility of 20cm$^2$ V$^{-1}$s$^{-1}$ which is very less as compared to the thin films prepared by molecular beam epitaxy (MBE) which has carrier mobility as high as 1330 cm$^2$V$^{-1}$s$^{-1}$ [32,37,39,40].

Table I. Calculated parameters from the Hall data and WAL fitting at T = 1.78K

| $N_{3D}$ ($10^{19}$ cm$^{-3}$) | µ(cm$^2$V$^{-1}$s$^{-1}$) | α | $B_\varphi$ | $l_\varphi$(nm) |
|---|---|---|---|---|
| 9 | 20 | 0.48 | 4.39e-2 | 57.23 |

Weak antilocalization (WAL) effect has been observed in most TI systems and is a hallmark of the presence of 2D surface states. Weak antilocalization (WAL) in TI systems has a twofold origin, time reversal symmetry and a π Berry phase acquired by the surface Dirac electrons going around the Dirac cone in momentum space due to spin momentum locking[29,30]. At low enough temperatures, where the phase coherence length is large compared to the grain size or diffusion length, quantum interference between different paths become important. In systems with time reversal symmetry, the wave functions corresponding to the two time-reversed scattering loops[20] of the electrons interfere constructively, thereby suppressing the conductivity, leading to weak localization (WL). Now in systems with strong spin-orbit coupling with a single Dirac cone, as the surface electron goes around the Dirac cone in momentum space, its spin orientation is rotated by 2π, and the wave function accumulates a π Berry phase owing to the two-component spinor wave function[21,41]. Hence, the wave functions corresponding to the forward and reverse paths in a loop interfere destructively, enhancing the conductivity. Thus the motion of electron is topologically protected. This is termed as weak anti-localization (WAL). It can be probed by means of applying small magnetic field perpendicular to the system which in turn disrupts time reversal symmetry of electrons in TI giving rise to positive magnetoresistance (MR). It must be noted here that WAL originating from the surface Dirac fermions is sensitive only to the perpendicular component of the applied magnetic field which has been experimentally verified by various groups in their angle dependent MR experiments on TI thin films or other nanostructures[22,29,30,42]. Origin of WAL in TI systems is completely different from those observed in 2D electron gases with spin orbit coupling. In 2D electron gases a little deviation from the random orientation of electron spin due to spin orbit coupling can give rise to WAL in the very low field regime but goes to WL in the higher field exhibiting an "M" like shape in MR[43]. This type of MR in 2D electron systems is not as robust as in TI systems since it persists only in very low temperature limits, while in TI it has been shown to persist even up to 100K[44]. Therefore the robustness to the electron-phonon interaction and non-saturating nature of MR in TI systems makes them completely different from the other trivial 2D electron gases.

This phenomenon is sensitive to different length/time scales in the system, such as phase coherence length ($l_\varphi$), elastic scattering length ($l_e$) and spin orbit scattering length ($l_{so}$). It is the competition between these different length/time scales which determine the ground state of these systems. For strongly spin orbit coupled systems like TIs, WAL is a characteristic feature of the surface states and spin orbit scattering rate dominates over the other two scattering rates, therefore can be ignored in the absence of magnetic impurities[30].

Magnetoconductance (MC) or Magnetoresistance (MR) data at different temperatures is shown in Fig. 4. The presence of a sharp cusp near zero magnetic fields in 1.78K, 2K, 3K, 7K and 25K data is quite evident while it vanishes at temperatures higher than 25K, where sharp cusp like behaviour near zero magnetic field diminishes, owing to reduction in phase coherence length due to the increased electron phonon interactions[23,30]. But there is no parabolic dependence of MC/MR even at temperatures as high as 40K as is also reported in some other works on different TI samples. It is clear from the figure that cusp like feature is persistent even at 25K showing unusual nature of MC/MR in these samples which indicates the presence of WAL at even higher temperature, owing to quantum interference of spin momentum locked Dirac surface states which make them topologically robust.

We have fitted all the data with the Hikami, Larkin and Nagaoka (HLN) quantum interference model for two dimensional systems in the strong spin orbit coupling limit as given in equation (1),

$$\delta G_{WAL}(B) \equiv G(B) - G(0) \cong \alpha \frac{e^2}{2\pi^2\hbar} \left[\Psi\left(\frac{1}{2} + \frac{B_\varphi}{B}\right) - \ln\left(\frac{B_\varphi}{B}\right)\right] \text{----------------------------} (1)$$

where, Ψ is digamma function and $B_\varphi = \hbar/4e\, l\varphi^2$ is characteristic magnetic field corresponding to the coherence length $l_\varphi$. The value of coefficient α signifies effective number of coherent channels contributing to conduction[45]. The MC data at 2K is fitted with the HLN equation and is shown in Fig. 4 (b).

WAL effect originating from the Dirac surface states ought to provide a value of α=0.5 in an ideal case[21,29,37]. In our case we have fitted the conductivity data in the entire magnetic field range 0 to ±8T and obtained an α of 0.40 at 2K, which is close to the values reported earlier for the sympletic case. From this fit it is very important to notice that HLN equation models the data very well up to 2T and thereafter deviates in the high field regime setting an upper field limit to fit the WAL data. This observation provides us important information about WAL behaviour in weak field regime in the TI system. On this basis we carried out HLN fitting of MC data up to 2T and for 1.78K as shown in the inset of Fig. 4 (b). Extracted values of prefactor α and phase coherence length from a fit to the low field data at 1.78K are provided in Table. I.

This value indicates that there is a dominant single channel coherent conduction in the system supporting the fact that WAL in BST thin films is 2D in nature. We carried out similar fitting for higher temperatures and extracted the values of the parameters α and $l_\varphi$. The values were in good agreement with the previous studies on different TI systems[23,29,42]. The exact reason for merely a single channel contributing to 2D WAL still remains a mystery. It has been speculated that due to the lattice mismatch between the substrate and bottom surface of the film, the bottom surface is highly disordered and $l_\varphi$ is significantly reduced in that case. But this explanation raises a question on the robustness and topological protection of the surface states even if it's the bottom surface which in principle cannot be localized even in the presence of heavy disorder unless the impurities are magnetic in nature which breaks the time reversal symmetry[3,46]. Therefore it cannot in principle be certified for thin films that only upper surface is contributing to conduction as a single channel. What we can possibly infer from the values of α is that there is a coherently coupled surface to bulk single and effective channel responsible for this type of conduction[30].

As it can be seen from the Fig. 4(a) the conductance from +2T to high magnetic limit does not saturate but shows a linear like trend right up to +8T. This linear MR displaying a non-saturating behaviour has been reported by many groups in other samples like BTS. Similar studies have been carried out by Assaf[44] and they have suggested possible mechanism for linear magnetoresistance in BTS thin films, to be the linear dispersion relation of the surface states or depleted bulk carriers. It can be clearly seen from the figure that linear magnetoresistance nearly becomes flat once temperature goes above 25K while below this temperature limit the sample exhibits linear MR behaviour suggesting that the effect is quantum mechanical in origin[47]. The linear MR regime is being explored in details and will be reported elsewhere.

To explore further the dimensionality and robustness of the system in terms of conduction we plotted the coherence length as a function of temperature. Fig. 5 (a) shows plot of extracted phase coherence length with temperature. Like other TI systems BST thin films are also seen to exhibit power law dependence of $l_\varphi$ on temperature ($l_\varphi \sim T^{-0.44}$) which is close to the 2D transport which ought to show a dependence, $l_\varphi = T^{-0.5}$[22,23]. Hence, power law dependence of coherence length on temperature again supports the fact that WAL indeed originates from topologically protected surface states. It is striking that even the thin films having low mobility (20 $cm^2V^{-1}s^{-1}$ at 1.78K) are showing 2D character in conduction supporting topological protection of surface states and the fact that the topology of the surface states are very robust[29].

Fig. 5(b) shows that the value of α does change with temperature but does not go above 1 at higher temperatures (25k), which can be accounted for more number of effective parallel channels contributing to conduction with increasing temperature arising from thermally excited bulk carriers, consistent with the previous results.

## IV. Conclusion

In conclusion, we have synthesized and characterized topological insulator candidate BST thin films by pulsed laser deposition technique. To the best of our knowledge BST thin films have been fabricated for the first time by PLD technique, earlier only single crystals have been studied. We repeated the observations for other BST films of similar thickness and were found to agree with each other. As prepared thin films of average thickness of 120nm show pronounced 2D weak antilocalization behaviour despite having poor mobility of the order of $20cm^2 V^{-1}s^{-1}$. But resistance vs. temperature profile of as prepared thin films shows metallic profile down to 15K and takes slight upturn thereafter, origin of which is unclear till now, but can generally be attributed to the freezing of the bulk carriers and other possible mechanisms. This R-T behaviour of BST thin films is completely different from those of BST single crystals which exhibit insulating behaviour in the whole temperature range. In addition we have also observed linear magnetoresistance in these thin films above 2T and origin of which has been attributed to the surface Dirac states exhibiting linear dispersion. The phase coherence length decays as a function of temperature with an exponent (β) of -0.44 which is very close to the predicted value of -0.5 emphasizing the fact that these are indeed 2D topological insulators and the robustness of the topological states persisted up to 25K. This 2D nature is also corroborated by the value of α (0.5) that was obtained from the fit of the low temperature MC data to the HLN equation. The present study on PLD grown thin films is a step towards viable technology as not only are they economically viable to make but also provides an ideal platform for integration with other materials and device fabrication possibly leading to low power spintronics and topological quantum computation.

### Acknowledgement

The Authors would like to thank Ministry of Human Resource Development (MHRD) for financial assistance. S.S. would like to thank U.G.C. for financial assistance. R.K.G would like to thank Bhavesh Patel for help in programming. C.M. would like to thank P.K. Panigrahi for discussions and for taking informal lectures on Topological Insulators.

**Figures**

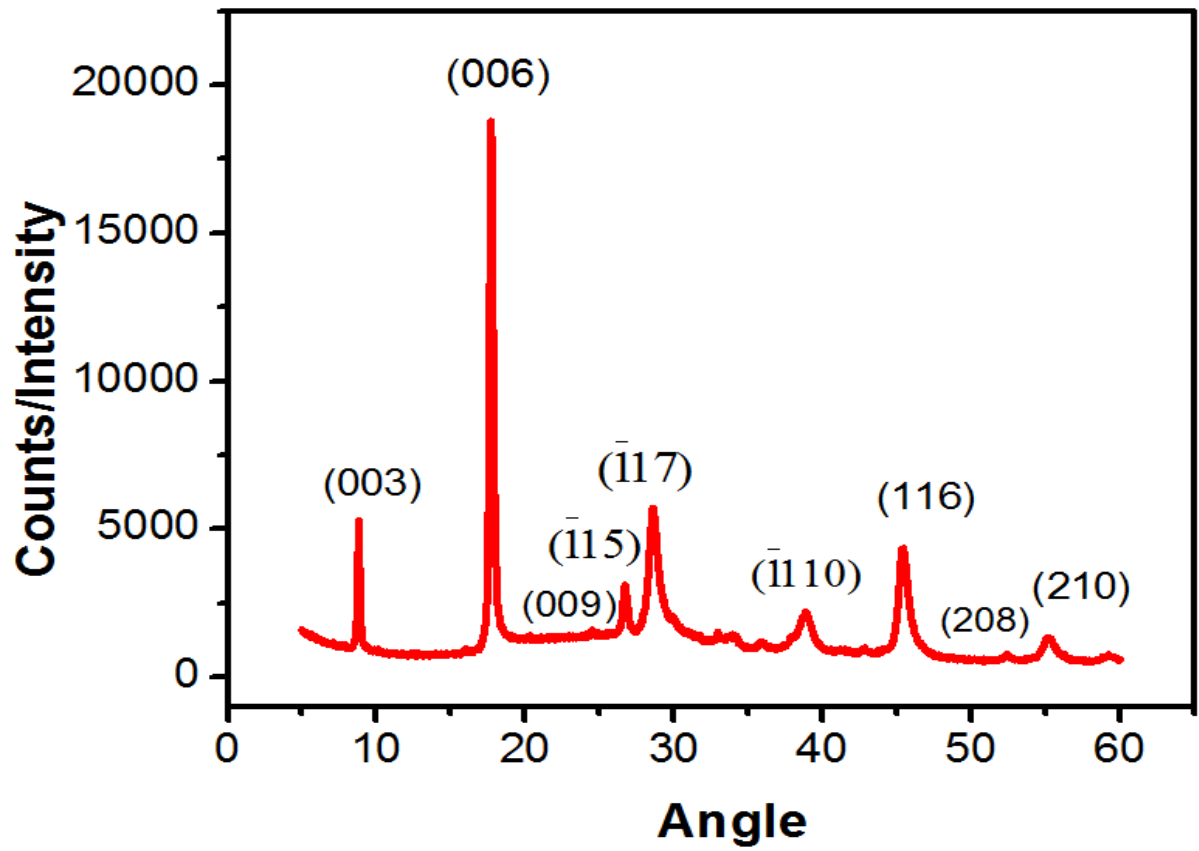

Fig.1 X- ray diffraction of as prepared thin films which shows (0, 0,3n) family of lines.

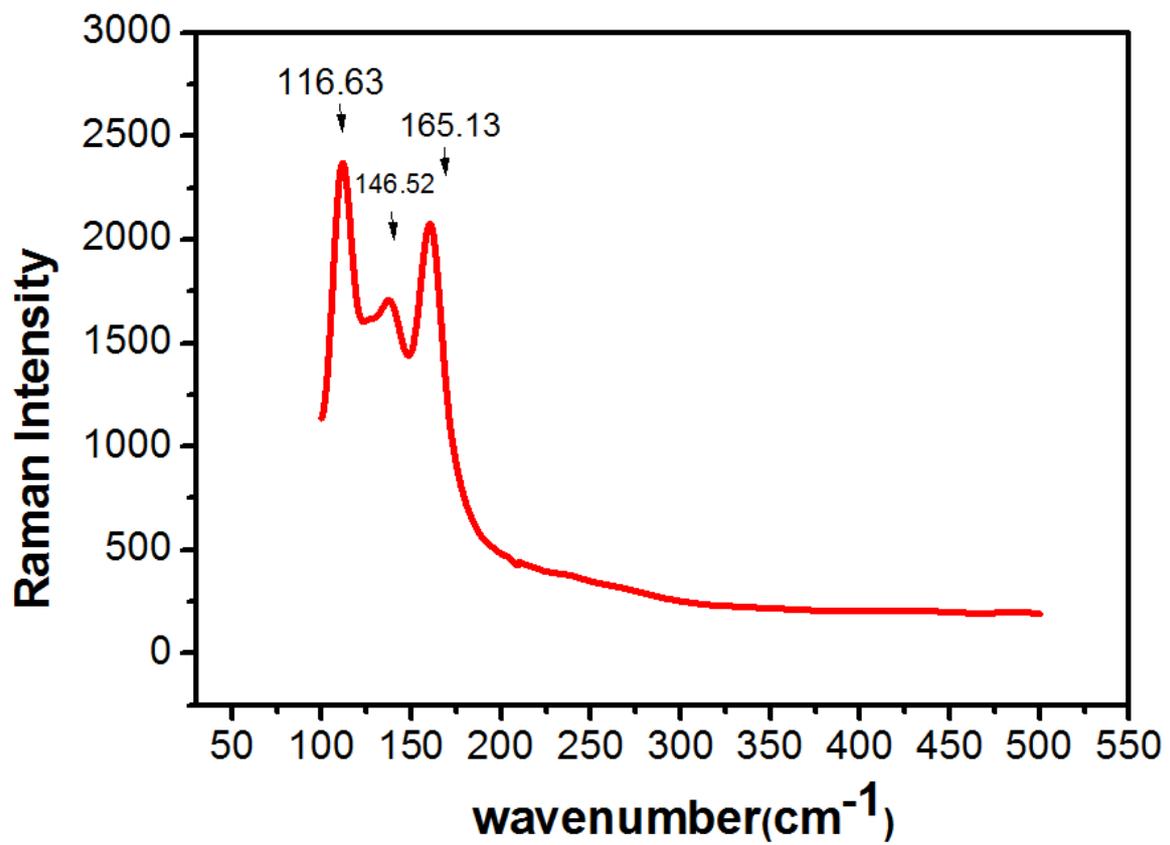

Fig. 2 Raman Spectrum of BST thin film. One can clearly see the three peaks at 116.6 cm$^{-1}$, 146.52 cm$^{-1}$ and 165.13 cm$^{-1}$ corresponding to BST.

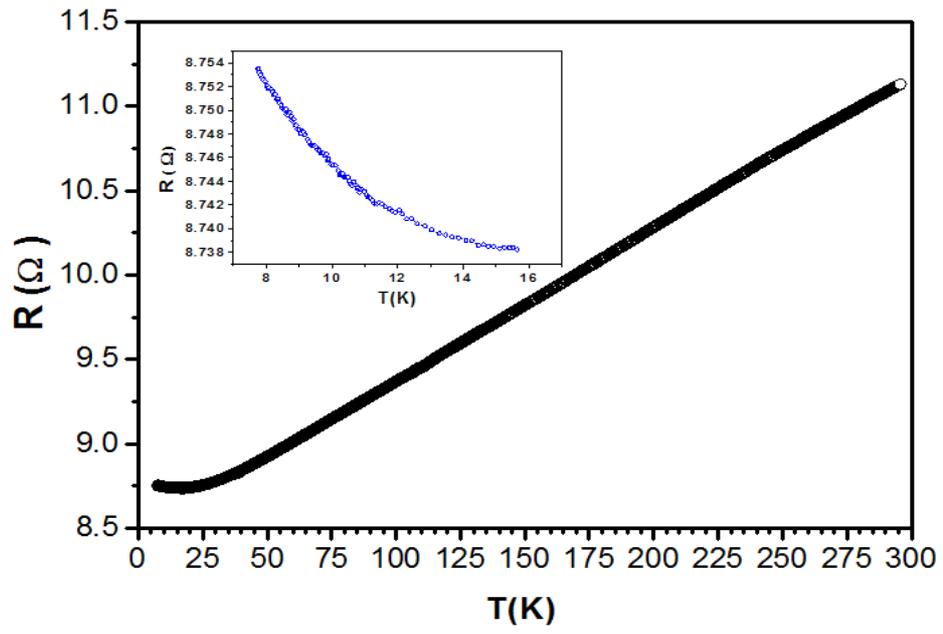

Fig. 3 Temperature dependence of resistance of BST thin films and the low temperature (below 15K) behaviour of R (inset).

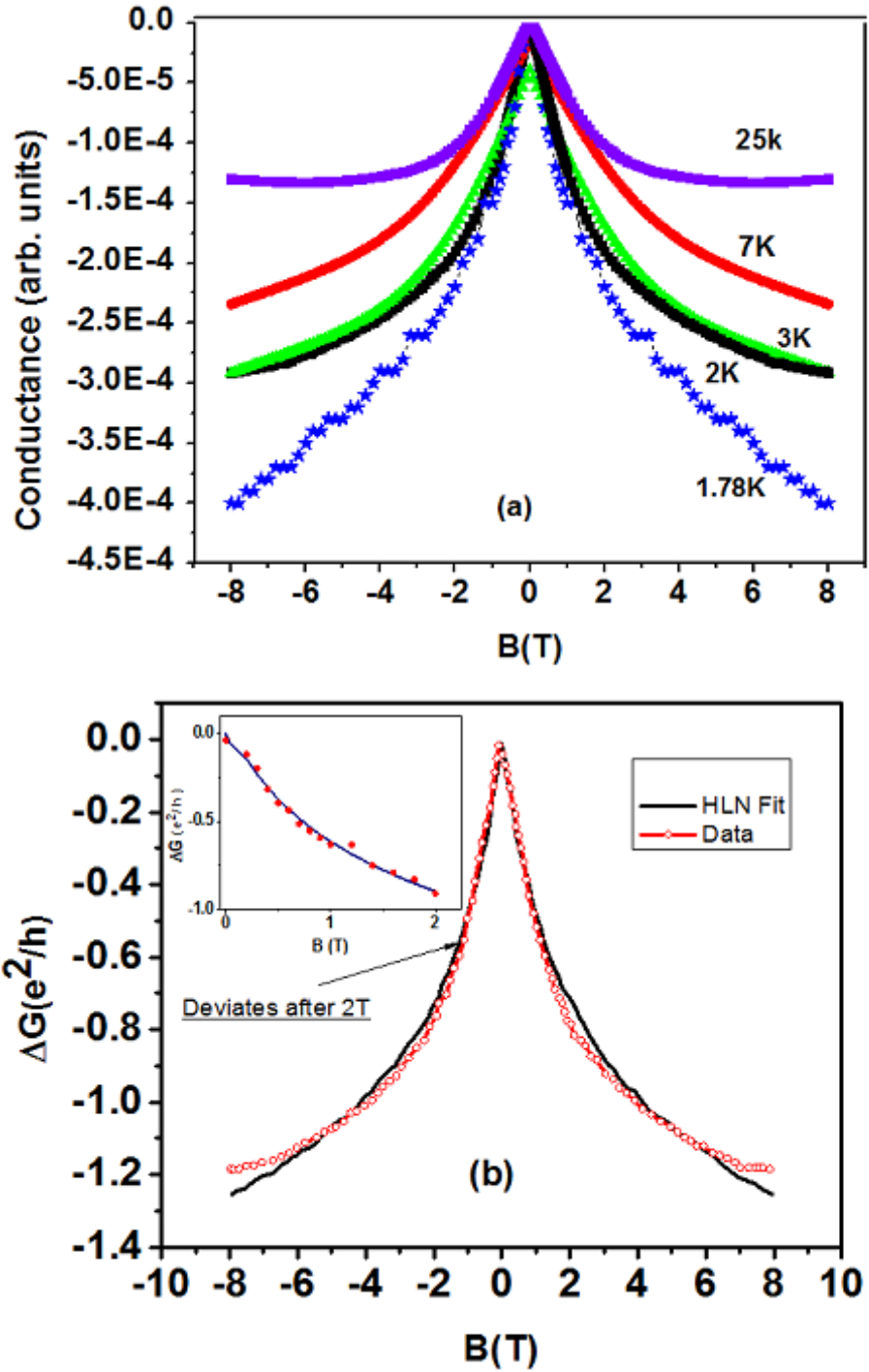

Fig. 4 (a) WAL at different temperatures 1.78K, 2K, 3K, 7K and 25K. At lower temperatures a sharp cusp is shown in low magnetic fields, which is a characteristic feature of WAL behaviour due to Dirac surface states. At higher temperatures, the cusp like nature of WAL curve diminishes due to reduction in phase coherence length. (b) Fit of the 2K data to the HLN model in the entire range. Low field curve fit is shown in the inset.

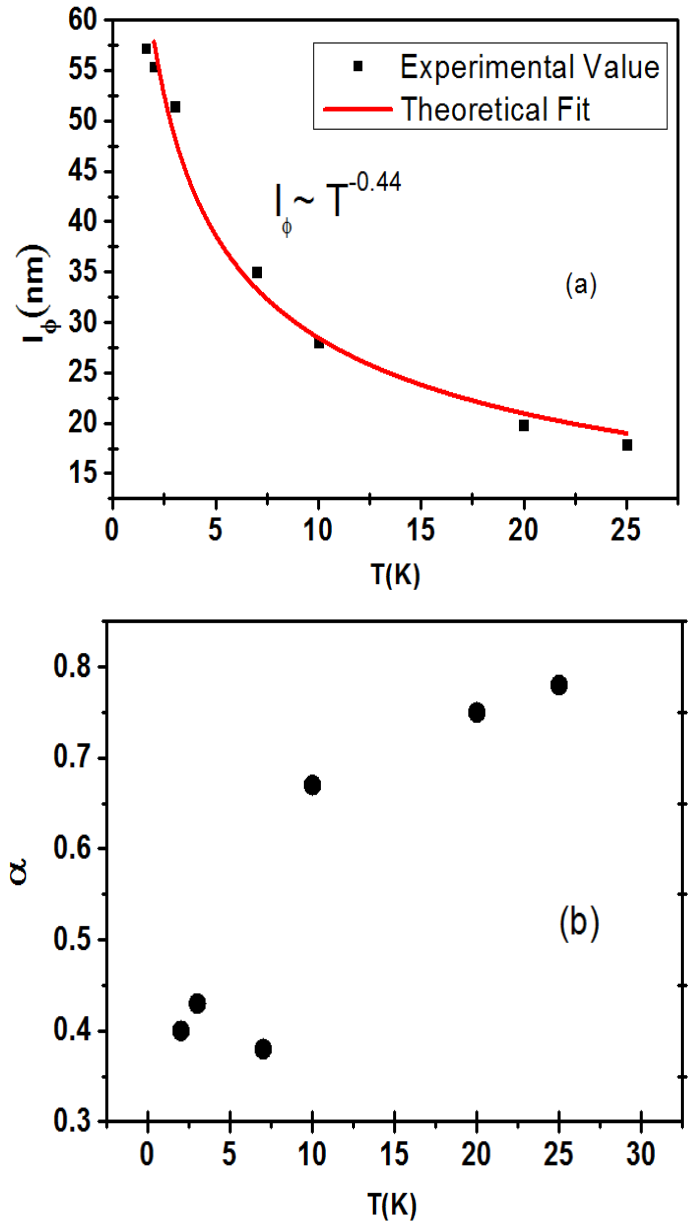

Fig. 5(a) Variation of $l_\varphi$ as a function of temperature T, fit to $l_\varphi \sim T^\beta$, yields $\beta=-0.44$ (b) Temperature dependence of $\alpha$.